\documentclass[reprint,amsmath,amssymb,aps,showkeys,floatfix,superscriptaddress]{revtex4-2}
\usepackage[utf8]{inputenc}
\usepackage[T1]{fontenc}
\usepackage[english]{babel}
\usepackage{graphicx}
\usepackage{siunitx}
\usepackage{amsmath,amssymb}
\usepackage{physics}
\usepackage{ulem}
\usepackage{lipsum}
\usepackage{hyperref}

\newcommand{\Mean}[1]{\langle #1 \rangle}
\newcommand{\Var}[1]{\mathrm{Var}\left[#1\right]}
\newcommand{\Cov}[2]{\mathrm{Cov}\left[#1,#2\right]}

\begin{document}

\title{A Simple and Robust Balanced Homodyne Detector for High‑Repetition‑Rate Pulsed Sources}

\author{Samuele Altilia}
\affiliation{Dipartimento di Fisica, Università degli Studi di Milano, Via Celoria 16, Milan, Italy}

\author{Edoardo Suerra}
\email{edoardo.suerra@unimi.it}
\affiliation{Dipartimento di Fisica, Università degli Studi di Milano, Via Celoria 16, Milan, Italy}

\author{Pietro Puppi}
\affiliation{Dipartimento di Fisica, Università degli Studi di Milano, Via Celoria 16, Milan, Italy}
\affiliation{Istituto Nazionale di Fisica Nucleare, Sezione di Milano, Via Celoria 16, Milan, Italy}

\author{Sebastiano Corli}
\affiliation{Dipartimento di Fisica, Università degli Studi di Milano, Via Celoria 16, Milan, Italy}

\author{Enrico Prati}
\affiliation{Dipartimento di Fisica, Università degli Studi di Milano, Via Celoria 16, Milan, Italy}
\affiliation{Istituto Nazionale di Fisica Nucleare, Sezione di Milano, Via Celoria 16, Milan, Italy}
\affiliation{Istituto di Fotonica e Nanotecnologie - CNR, Piazza Leonardo da Vinci 32, Milan, Italy}

\author{Simone Cialdi}
\affiliation{Dipartimento di Fisica, Università degli Studi di Milano, Via Celoria 16, Milan, Italy}
\affiliation{Istituto Nazionale di Fisica Nucleare, Sezione di Milano, Via Celoria 16, Milan, Italy}

\begin{abstract}
We  design and experimentally characterize a balanced homodyne detector optimized for high-repetition-rate (\SI{100}{\mega\hertz})  pulsed optical sources. Unlike conventional transimpedance-amplifier architectures, which suffer from nonlinearities and dynamic instabilities with ultrashort pulses, our approach allows to directly amplify the photocurrent extracted at the common photodiode node without feedback loops. A theoretical model describing the detector response, noise, and pulse-to-pulse correlations is developed, providing quantitative predictions for the signal variance, signal-to-noise ratio (SNR), and inter-pulse correlations. Implemented with two matched InGaAs photodiodes illuminated by a \SI{1030}{\nano\meter} mode-locked laser at \SI{100}{\mega\hertz}, the detector exhibits excellent linearity and shot-noise-limited scaling of the signal variance with optical power. Optimizing the temporal integration window yields a maximum SNR of about \SI{14}{\deci\bel}, while correlation measurements confirm negligible inter-pulse correlations. These results demonstrate that the proposed architecture offers a robust and simple solution for high-speed pulsed homodyne detection, suitable for quantum optics and continuous-variable quantum information applications.
\end{abstract}

\maketitle

\section{Introduction}

Balanced homodyne detection is a fundamental technique in quantum optics and advanced photonics \cite{Schnabel2017}, which enables direct measurement of the quadratures of the electromagnetic field \cite{Yuen1983,Leonhardt1995,Bachor2004}, making state tomography \cite{Lvovsky2009}, non-classical states analysis \cite{Lvovsky2015,Olivares2021}, quantum random number generation \cite{Gabriel2010}, and information processing in continuous-variable protocols possible \cite{Braunstein2005,Weedbrook2012}.

However, when moving from the continuous-wave regime to the pulsed regime, the situation changes drastically, as ultrashort optical pulses generate very intense current spikes and extremely fast edges, forcing the detector electronics to operate under much more severe conditions \cite{Hansen2001,Zavatta2002}. In these circumstances, the current-to-voltage conversion and the amplification stages become the limiting elements, and a large part of the literature shows how stability, linearity, and bandwidth become difficult to maintain \cite{Cooper2013,Guay2021}, especially in detectors based on transimpedance amplifiers (TIAs).

A first significant example is provided by Cooper et al.~\cite{Cooper2013}. In this work the authors achieve high performance in terms of SNR and stability, but the circuit requires extremely delicate design: detailed SPICE modeling, careful choice of compensation capacitances, and strict control of layout and parasitics. All these aspects clearly indicate that the OPA847 used in the design operates close to its dynamic limits. Ultrashort optical pulses of about\SI{100}{\femto\second} produce current spikes at the photodiode output that can push the feedback op-amp beyond its slew-rate limits, leading to saturation, overshoot, or oscillations.

A direct confirmation of these difficulties is provided by Guay and Genest~\cite{Guay2021}. They experimentally demonstrate that TIA-based detectors employing OPA847 operational amplifiers exhibit strong nonlinearities in the pulsed regime even when the average power is low. The authors observe dynamic saturation, pulse broadening, timing shifts, ringing, and sometimes chaotic behavior. The decisive evidence is that these distortions do not originate from the photodiodes, as shown by the fact that the nonlinearity appears only after the subtraction of the currents generated by the two photodiodes. It is therefore the operational amplifier in the TIA configuration, exposed to extreme pulsed currents, that leaves its safe operating region, exceeding slew-rate and phase-margin limits. This work makes it clear that, in the pulsed regime, conventional TIAs represent the primary obstacle to linearity and measurement repeatability.

An opposite approach is described by Hansen et al.~\cite{Hansen2001}. In that case no TIA is present: detection relies on a charge-sensitive preamplifier (Amptek A250) followed by A275 shaping amplifiers. The absence of an operational amplifier that must directly sink the photodiode current completely avoids slew-rate limits, oscillations, and pulsed saturation. The authors show excellent linearity and no pulse distortion. However, the system is very slow: the electrical pulses last almost a microsecond and the bandwidth is about 1~MHz. This scheme, although highly linear, therefore cannot operate at repetition rates on the order of 100~MHz.

Particularly interesting is also the scheme adopted by Zavatta et al.~\cite{Zavatta2002}. Here the photodiodes are not connected to the op-amp input through a TIA; instead, they dump the generated current into a load resistor. The RF operational amplifier (Comlinear CLC425) amplifies only the voltage developed across the resistor, without transimpedance feedback. This completely avoids the dynamic problems of TIAs: no pulsed saturation, no need to follow picosecond edges, and no critical loop to stabilize. The result is a detector that remains stable and linear up to \SI{82}{\mega\hertz}, despite the presence of an operational amplifier. It is an important example: what causes the issues is not so much the presence of an op-amp, but its use in a transimpedance configuration.

The most recent article, by Kouadou et al.~\cite{Kouadou2025}, directly tackles the 150~MHz limit and explicitly recognizes that the main challenge is the current-to-voltage conversion performed by the TIA. To achieve bandwidths between 60 and 110~MHz and pulse-by-pulse measurements up to 150~MHz, the authors design a TIA based on the OPA856, an operational amplifier intended for ultrawideband applications. Circuit stability depends critically on the compensation of the total loop capacitances (photodiode, amplifier input, and feedback capacitance). Incorrect tuning introduces gain peaking in the Bode plot and therefore instability or saturation, while excessive compensation reduces the useful bandwidth. The TIA works, but only thanks to extremely refined electronic design, operating at the edge of stability and entirely devoted to mitigating the intrinsic weaknesses of the transimpedance architecture.

Taken together, these results clearly show that the current-to-voltage conversion is the critical node of pulsed homodyne detection. When entrusted to a conventional TIA, it introduces nonlinearities and instabilities; when removed (as in~\cite{Hansen2001}) perfect linearity is achieved but bandwidth is lost; when replaced by a load resistor (as in~\cite{Zavatta2002}) linearity is preserved while relatively high repetition rates can be reached; when one instead wishes to keep the TIA at very high repetition rates (as in~\cite{Kouadou2025}), extremely sophisticated design and highly specialized amplifiers become necessary.

In this context we propose a scheme that aims to reconcile linearity, robustness, and speed, while avoiding the structural difficulties of op-amp TIAs. The idea is to use two balanced photodiodes that discharge onto a common load resistor, generating a voltage proportional to the current difference. This voltage is then amplified by a transistor stage, also free of critical feedback loops. In this way, no operational amplifier is forced to follow large pulsed transitions, the risk of dynamic saturation and oscillation is drastically reduced, and linearity is preserved while maintaining sufficient bandwidth for repetition rates on the order of 100~MHz. Moreover, the simplicity of the circuit makes it possible to describe its behavior accurately with a simple mathematical model.

\section{Theory}

In the standard homodyne configuration, the quantum field under investigation is interfered with a high-power local oscillator (LO) on a beam splitter. The two output ports are detected by a pair of photodiodes with high quantum efficiency (QE), and the homodyne signal is obtained by taking the difference between the two photocurrents. This technique effectively suppresses intensity noise that is common to the LO and selectively accesses the field quadrature in phase with the LO.

\begin{figure}
    \centering
    \includegraphics[width=0.4\linewidth]{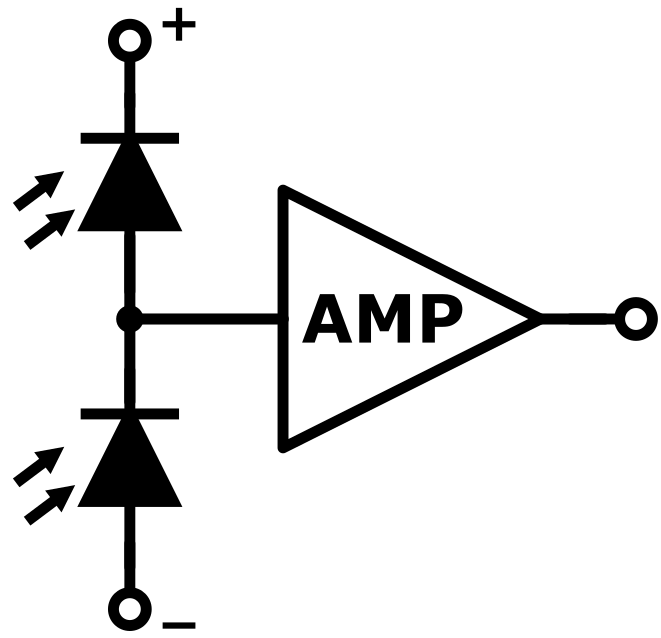}
    \caption{Principle of operation of the balanced detector, where the difference photocurrent is extracted at the common node of the two photodiodes and subsequently amplified.}
    \label{fig:detector_simple}
\end{figure}

In this work, we focus on the case where both the signal and the LO consist of ultrashort pulses with a high repetition rate in the hundreds of \si{\mega\hertz} range, as is typical for mode-locked sources. In this regime, the scheme shown in Fig.~\ref{fig:detector_simple} is commonly adopted, in which the difference photocurrent is extracted directly at the common node of the two photodiodes and subsequently amplified.

We now derive the general detection theory, for which entering circuit-level details is not required, relying only on the sole assumption, which is well satisfied in our implementation, that the detector response is \textit{linear} with respect to the two incident optical signals. Denoting by $x_{1,2}(t)$ the optical signals impinging on the two photodiodes and by $y(t)$ the detector output, linearity requires
\begin{equation}
    y(t) = \int \big[ h_1(\tau)\, x_1(t-\tau) - h_2(\tau)\, x_2(t-\tau) \big] \,d\tau + r(t) \,,
    \label{eq:linearity}
\end{equation}
where $h_{1,2}(t)$ are the respective impulse response functions, and $r(t)$ models the electronic noise at the output.

To simplify the discussion, we temporarily disregard the pulse train and consider a single ultrashort pulse. Being the pulse duration much shorter than the response time of the photodiodes, the two optical signals can be modeled as $x_{1,2}(t) = n_{1,2}, \delta(t - t_{1,2})$, where $n_{1,2}$ is the number of photons absorbed by each photodiode and $t_{1,2}$ are the corresponding pulse arrival times. Substituting into Eq.~\ref{eq:linearity} simply yields
\begin{equation}
    y(t) = n_1 h_1(t - t_1) - n_2 h_2(t - t_2) + r(t) \,.
\end{equation}
From shot to shot, $y(t)$ is a stochastic variable due to the randomness of both $r(t)$ and $n_{1,2}$. Since the electronic noise is zero-mean and uncorrelated with the optical fields, \mbox{$\Mean{r(t) \, n_{1,2}} = 0$}, we can write the mean and the variance of $y(t)$ as
\begin{align}
    \Mean{y(t)} &= \Mean{n_1} h_1(t-t_1) - \Mean{n_2} h_2(t-t_2) \,, \label{eq:y_mean} \\
    \notag \\
    \Var{y(t)} &= \Var{n_1} h_1^2(t-t_1)
        + \Var{n_2} h_2^2(t-t_2) + \notag \\
        & - 2 \, \Cov{n_1}{n_2} \, h_1(t-t_1) \, h_2(t-t_2)
        \, + \notag \\ &+ \Var{r(t)} \,.
    \label{eq:y_var}
\end{align}

The number of photons absorbed by the two photodiodes fluctuates due to two contributions: quantum noise and classical laser noise, which we represent as two zero-mean stochastic variables, $\delta n^{(\mathrm{q})}_{1,2}$ and $\delta n^{(\mathrm{c})}_{1,2}$, respectively. 
We thus write
\begin{equation}
    n_{1,2} = \Mean{n_{1,2}} + \delta n^{(\mathrm{q})}_{1,2} + \delta n^{(\mathrm{c})}_{1,2} \,.
\end{equation}

When the homodyne input is vacuum, as in our detector-testing configuration, the quantum noise is purely shot noise and its variance equals the mean photon number, \mbox{$\Var{\delta n^{(\mathrm{q})}_{1,2}} = \Mean{n_{1,2}}$}.
The classical laser noise, by contrast, consists of intensity fluctuations proportional to the mean intensity itself and is typically modeled in terms of a \textit{relative intensity noise} (RIN) coefficient, here denoted as $\kappa$. Its variance is therefore \mbox{$\Var{\delta n^{(\mathrm{c})}}_{1,2} = \kappa^2 \Mean{n_{1,2}}^2$}.
Furthermore, the shot-noise contributions in the two outputs are uncorrelated with each other and uncorrelated with classical noise, \mbox{$\Mean{ \delta n_1^{(\textrm{q})} \delta n_2^{(\textrm{q})} } = \Mean{ \delta n^{(\textrm{q})} \delta n^{(\textrm{c})} } = 0$}, whereas classical noise is common to both paths, so that \mbox{$\Mean{ \delta n_1^{(\textrm{c})} \delta n_2^{(\textrm{c})} } = \kappa^2 \Mean{n_1} \Mean{n_2}$}. Using these relations, the variance in Eq.~\ref{eq:y_var} becomes
\begin{equation}
\begin{aligned}
    \Var{y(t)} &= \Mean{n_1} h_1^2(t-t_1) + \Mean{n_2} h_2^2(t-t_2) \, + \notag \\
    & + \kappa^2 \left[ \Mean{n_1} h_1(t-t_1) -  \Mean{n_2} h_2(t-t_2) \right]^2 + \notag \\ &+ \Var{r(t)} \,.
    \label{eq:y_varianza}
\end{aligned}
\end{equation}
This expression explicitly shows that accessing the shot-noise-limited performance requires not only low electronic noise but also excellent balance between the two detection channels: the optical powers must be well matched and so the two impulse responses and pulse arrival times. Any imbalance introduces noise with a \textit{quadratic} dependence on optical power, rather than the \textit{linear} one characteristic of shot noise. When this match is well met, we get
\begin{equation}
    \Var{y(t)} = 2 \Mean{n} h^2(t) + \Var{r(t)} \,, 
    \label{eq:y_varianza_bilanciata}
\end{equation}
where the response functions and the mean photon numbers are now equal for the two photodiodes and we have set $t=0$ at the two equal arrival times. In this case the variance of the signal, apart from a background contribution due to electronic noise, reproduces the square of the response function $h$ as the considered time instant is varied.

Having established this behavior, we now turn to the reconstruction of the actual homodyne signal. To this end, the detector output must be integrated over a time window $\Delta t$ around the expected arrival time of the pulses, so as to collect the entire charge produced by the difference photocurrent: we ask therefore how this integration window should be chosen. We start by writing the integrated signal
\begin{equation}
    Y(\Delta t) = \int_{\Delta t} y(t) \, dt = n_1 H_1(\Delta t) - n_2 H_2(\Delta t) + R(\Delta t),
\end{equation}
where we have defined $H_{1,2}(\Delta t) = \int_{\Delta t} h_{1,2}(t - t_{1,2})\, dt$ and $R(\Delta t) = \int_{\Delta t} r(t)\, dt$.  
Its mean and variance are
\begin{align}
    \Mean{Y(\Delta t)} &= \Mean{n_1} H_1(\Delta t) - \Mean{n_2} H_2(\Delta t), \\
    \notag \\
    \Var{Y(\Delta t)} &= \Var{n_1} H_1^2(\Delta t) + \Var{n_2} H_2^2(\Delta t) \, + \notag \\                  
    & - 2 \, \Cov{n_1}{n_2} H_1(\Delta t) \, H_2(\Delta t) \, + \notag \\ &+ \Var{R(\Delta t)} \,,
\end{align}
and under the same assumptions made above we get
\begin{equation}
    \Var{Y(\Delta t)} = 2 \Mean{n} H^2(\Delta t)  + \Var{R(\Delta t)} \,.
\end{equation}
We can therefore define a signal-to-noise ratio (SNR) in terms of the variance of the integrated signal measured with and without the optical input, thereby quantifying how the detector’s electronic noise degrades the homodyne measurement: 
\begin{equation}
    \mathrm{SNR}^2(\Delta t) = \frac{\Var{Y(\Delta t)}}{\Var{R(\Delta t)}} = 1 + \frac{ 2\Mean{n} H^2(\Delta t)}{\Var{R(\Delta t)}} \, .
\label{eq:SNR}
\end{equation}
This function exhibits a maximum as a function of $\Delta t$, corresponding to the optimal trade-off between capturing the full photogenerated charge and avoiding the accumulation of low-frequency electronic noise. Since both the integrated response function and the variance of the electronic noise can be readily measured, a direct comparison between the prediction of Eq.~\ref{eq:SNR} and the experimentally measured values of the SNR is possible.

Finally, we consider the correlations between different pulses of the homodyne signal. Measuring these correlations is important to ensure that the detector is indeed capable of discriminating between different pulses.
To this end, we introduce the optical signals of the pulse train
\begin{equation}
    x_{1,2}(t) = \sum_j n_{1,2;j} \, \delta(t-jT-t_{1,2}) \,,
\end{equation}
where $n_{1,2;j}$ is the number of photons absorbed by each photodiode due to the $j$-th pulse, and $T = 1/f_{\textrm{rep}}$ is the inverse of the source repetition rate. We now assume, as before, perfect balancing of the detectors. From Eq.~\ref{eq:linearity} we then obtain
\begin{equation}
    y(t) = \sum_j (n_{1;j} - n_{2;j}) h(t-jT) + r(t) \,,
\end{equation}
which, when evaluated at times that are multiples of $T$, can be written as
\begin{equation}
    y_k = \sum_j (n_{1;j} - n_{2;j}) h_{k-j} + r_k \,,
\end{equation}
where we have defined $y_k = y(kT)$, and similarly for $h$ and $r$.
Note that the case of practical interest is when the origin of time is chosen such that the function $h$ reaches its peak there, so that the signal variance is maximal.

We can now define the correlator between different pulses as
\begin{equation}
    C_k = \frac{ \Mean{y_j y_{j+k}} - \Mean{y_j}\Mean{y_{j+k}} }{\sqrt{\Var{y_j}} \sqrt{\Var{y_{j+k}}}} = \frac{ \Mean{y_j y_{j+k}} }{ \Mean{y_j^2} } \,,
\end{equation}
where in the last step we used that $y_k$ is stationary and zero mean.

If vacuum enters the homodyne together with the LO, different pulses are uncorrelated and we obtain
\begin{equation}
    C_k = \frac{ 2\Mean{n}\sum_l h_{l} h_{l+k} + \Mean{r_j r_{j+k}} }{ 2\Mean{n} \sum_l h_{l}^2 + \Mean{r_j^2 } } \,.
\end{equation}

In the relevant case where the electronic noise is negligible compared to the homodyne signal, we finally get
\begin{equation}
    C_k = \frac{ \sum_l h_{l} h_{l+k} }{ \sum_l h_{l}^2 } \,.
\label{eq:corr_approx}
\end{equation}

From this expression we see that, for the correlation to vanish, the response function $h$ must decay within a single period of the source. In practice, however, this limitation is not set solely by the detector speed -which may be sufficient if the bandwidth is large enough - but by the need to filter the output signal in order to suppress residual laser noise that cannot be fully removed by detector balancing alone, due to non-idealities and asymmetries in the system. These filters must be chosen to be spectrally very narrow - something more easily achieved through digital filtering - so that the correlations introduced between different pulses remain limited.

\section{Experimental realization}

\subsection{Detector scheme}

\begin{figure}
\centering
\includegraphics[width=\linewidth]{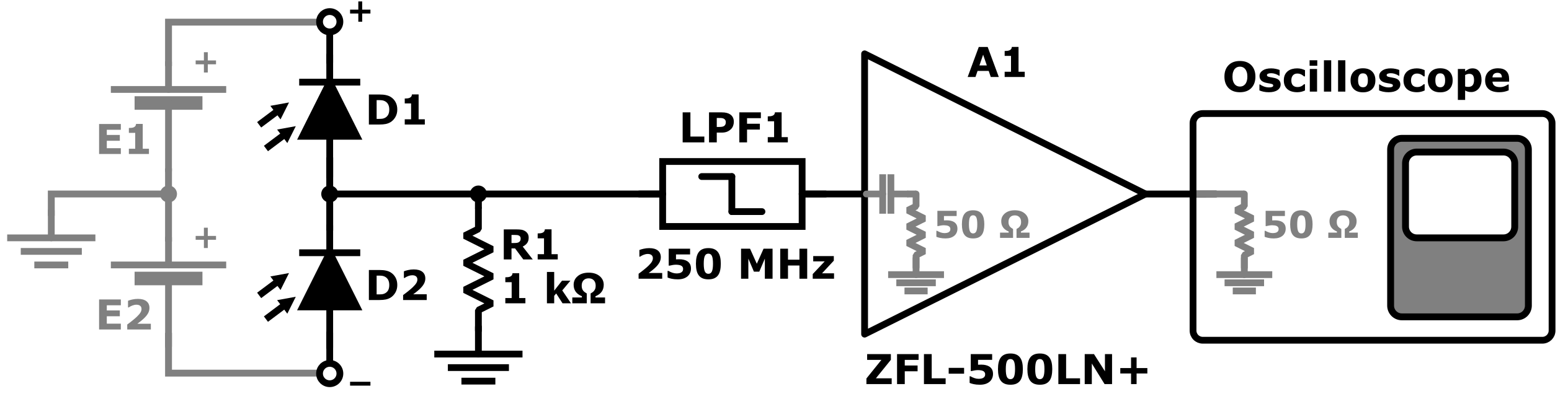}
\caption{Electronic scheme of the homodyne detector. Two InGaAs photodiodes (D1, D2) generate the difference photocurrent at their common node, which is filtered (LPF1), amplified (A1), and sent to an oscilloscope for acquisition. The bias voltages (E1, E2) and resistor $R_1$ set the operating point and ensure stable detector response.}
\label{fig:detector_scheme}
\end{figure}

The electronic scheme of our homodyne detector is shown in Fig.~\ref{fig:detector_scheme}. The two photodiodes D1 and D2 (InGaAs Fermionics FD500) were carefully selected to ensure closely matched quantum efficiencies. Their bias voltages, properly filtered, can be finely adjusted and are represented here by the voltage sources E1 and E2.

The common node between D1 and D2 is connected to the \SI{250}{\mega\hertz} low-pass filter LPF1, which reduces the pulse peak amplitude while preserving sufficient bandwidth to temporally resolve them. Since the noise introduced by the oscilloscope depends on the selected scale, this filtering stage is necessary to allow the use of a sufficiently small vertical scale, thereby avoiding degradation of the SNR. The signal is then fed into the low-noise commercial amplifier A1 (Mini-Circuits ZFL-500LN+). Since the amplifier is capacitively coupled, the resistor R1 is required to set the operating point of the photodiodes. Its value (\SI{1}{\kilo\ohm}) must be much larger than the filter input impedance (\SI{50}{\ohm}) to avoid significant signal loss, yet small enough to prevent shifts in the photodiode operating point if the optical power of the LO is varied, due to small current imbalances. 

The amplified signal is finally fed directly into the \SI{50}{\ohm} input of an oscilloscope (Rohde \& Schwarz MXO58) for data acquisition and analysis.

\subsection{Optical setup}

\begin{figure}
\centering
\includegraphics[width=\linewidth]{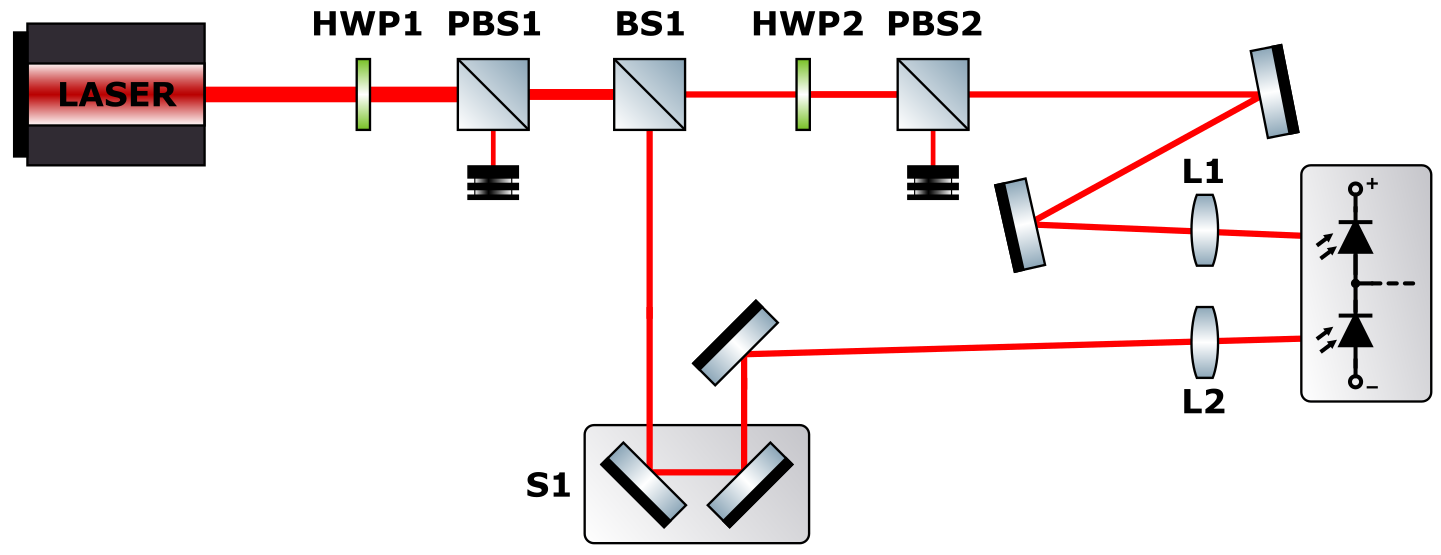}
\caption{Homodyne optical setup used to test the detector. The LO from a mode-locked laser is power-adjusted via HWP1 and PBS1, mixed with vacuum at BS1, and focused onto the photodiodes by L1 and L2. Fine power and path-length adjustments are made with HWP2 and PBS2, and the translation stage S1.}
\label{fig:optical_setup}
\end{figure}

Figure~\ref{fig:optical_setup} shows the schematic of the homodyne optical setup we used to test the detector. Laser pulses are generated by a mode-locked source at a central wavelength of $\lambda = \SI{1030}{\nano\meter}$, with a repetition rate of $f_\textrm{rep} = \SI{100}{\mega\hertz}$, which serves as the LO of the homodyne detector. The LO power can be adjusted by means of the half-wave plate HWP1 and the polarizing beam splitter PBS1, which act as an amplitude modulator.

The LO is then directed to the homodyne beam splitter BS1, whose second input port is left in the vacuum state. The beam splitter BS1 is slightly unbalanced, thus an additional amplitude modulator stage composed of HWP2 and PBS2 is placed on the higher-power output arm to allow for fine power balancing. On the other output arm, a translation stage S1 is used for fine adjustment of the optical path length.

The optical path length is matched interferometrically by directing the two beams which go towards the detector at a small relative angle and observing the formation of interference fringes at their intersection point. In this way, the path-length matching is performed at the interferometric level, and the residual error is dominated by the mechanical positioning of the detector, which is nevertheless negligible.

The two beams reaching the detector are focused onto the photodiodes by lenses L1 and L2, each with a focal length of \SI{150}{\milli\meter}. This produces a beam waist of approximately \SI{50}{\micro\meter}, much smaller than the photodiode diameter (\SI{500}{\micro\meter}) to ensure full collection of the optical beam, yet large enough to avoid quantum efficiency fluctuations caused by position-dependent variations on the photodiode surface due to mechanical vibrations of the alignment system.

\section{Measurements}

\subsection{Detector response linearity}

As discussed above, our homodyne detector model assumes a linear response to the incident optical signal. As a first step, we therefore validated the linearity of the photodiode response function by confirming the invariance of its temporal shape and the linear scaling of the peak voltage with the optical power.

To this end, we performed measurements at various average powers of the laser pulses, $P = 1 \divisionsymbol \SI{5}{\milli\watt}$, illuminating one photodiode at a time (\SI{5}{\milli\watt} corresponds to a photocurrent close to the maximum permissible reverse current of \SI{5}{\milli\ampere}). In this completely unbalanced configuration, the resulting large differential signal would drive the amplifier into saturation, thus it could not be used. Instead, we connected the output of LPF1 directly to the oscilloscope, set to an input impedance of $\SI{50}{\ohm}$. Since the amplifier exhibits a highly linear response when operated within its specified range, the linearity verified in this configuration is preserved also in the final balanced setup, in which the amplifier is reintroduced and operated within its linear regime.

In the following, we write the voltage signal measured on the oscilloscope in response to a single laser pulse as
\begin{equation}
    \tilde{y}(t) = \tilde{V}_p \, h_{\textrm{norm}}(t) \,,
\end{equation}
where the tilde indicates signals measured without the amplifier, $\tilde{V}_p$ is the peak voltage, and the subscript "norm" denotes a response function normalized to unity at its peak.

\begin{figure}
\centering
\includegraphics[width=0.9\linewidth]{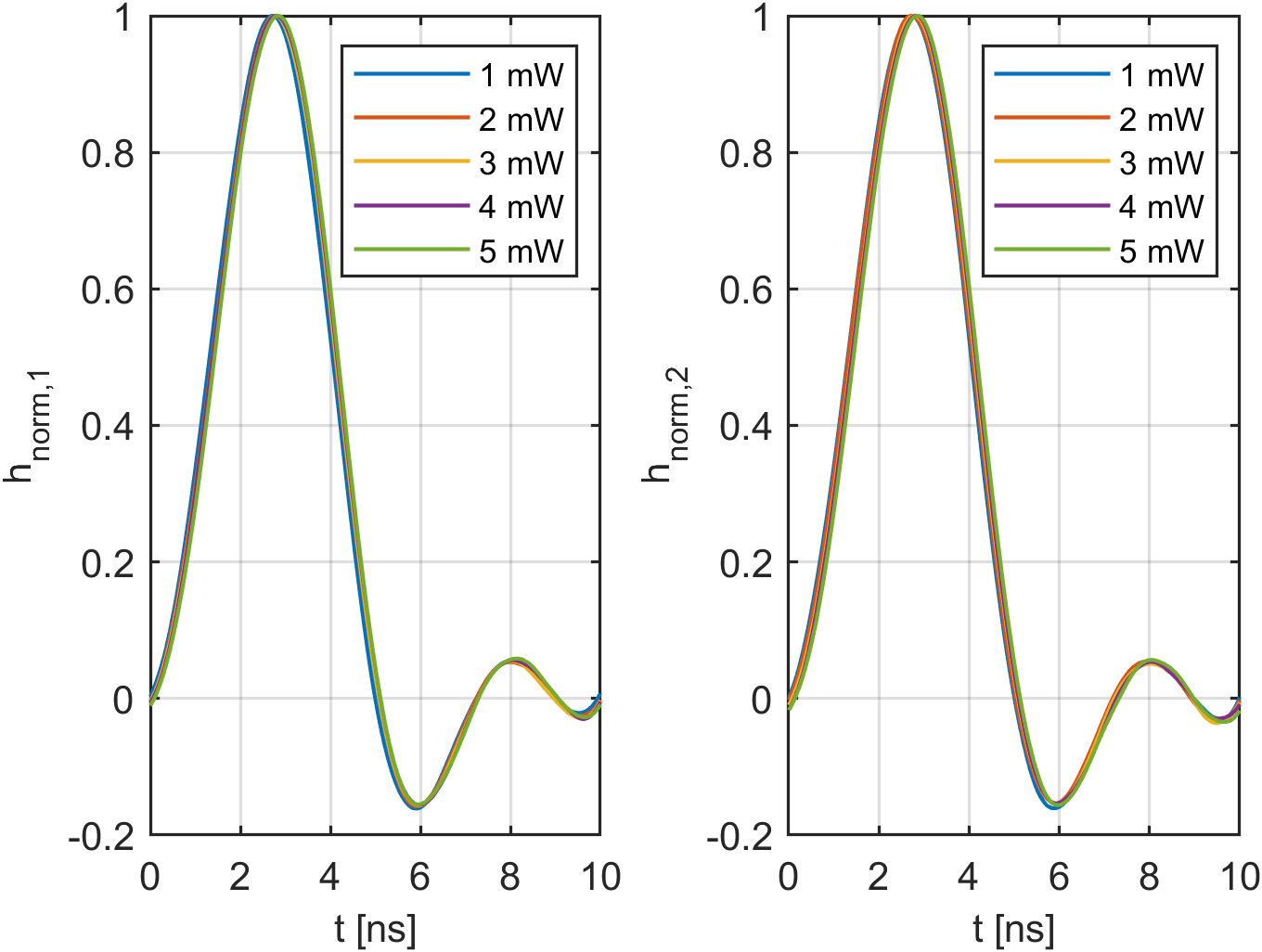}
\caption{Invariance of the peak-normalized response functions $h_{\mathrm{norm}}$ of the two photodiodes at the output of LPF1 as a function of the average optical power $P$.}
\label{fig:h1_h2_shape_invariance}
\end{figure}

Figure~\ref{fig:h1_h2_shape_invariance} shows the normalized response functions $h_{\textrm{norm}}$ for both photodiodes at the different tested optical powers, demonstrating that its shape remains essentially unchanged.

\begin{figure}
\centering
\includegraphics[width=0.8\linewidth]{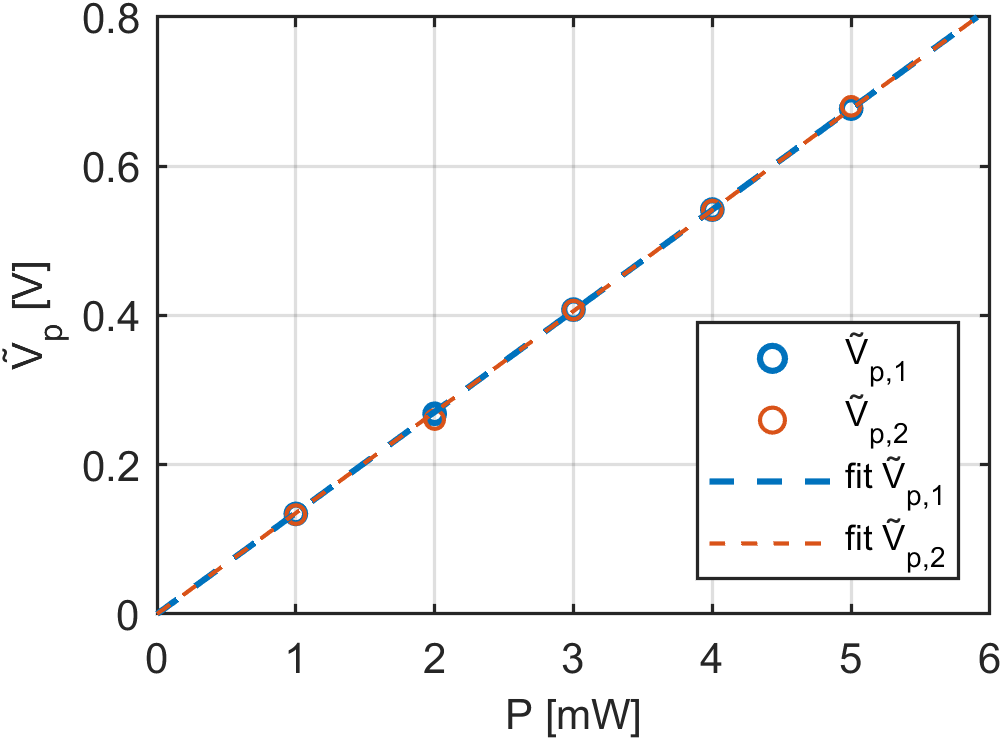}
\caption{Linearity of $\tilde{V}_p$ as a function of $P$ for the two photodiodes, demonstrating the absence of saturation effects. The slopes obtained are $K_1 = 0.1354 \pm 0.0002,\si{\volt\per\milli\watt}$ and $K_2 = 0.1353 \pm 0.0007,\si{\volt\per\milli\watt}$; the coefficients of determination are $R_1^2 = 0.9999$ and $R_2^2 = 0.9994$.}
\label{fig:h1_h2_linearity}
\end{figure}

The linearity of $\tilde{V}_p$ as a function of $P$ is instead shown in Fig.~\ref{fig:h1_h2_linearity}. To determine the expected slope $K$ of this linear dependence, we integrate the measured voltage signal over one period of the source, thereby yielding the total charge $Q$ released during a single pulse. Since, aside from the purely reactive low-pass filter, the detector signal discharges across the oscilloscope input resistance $R = \SI{50}{\ohm}$, we can therefore write
\begin{equation}
    \int_T \tilde{y}(t)\, dt
    = \tilde{V}_p \int_T h_{\textrm{norm}}(t)\, dt
    = R\, Q
    = \frac{R\, \eta \, q\, P\, T}{h \nu},
\end{equation}
where $q$ is the elementary charge, $h$ is Planck’s constant, $\nu = c/\lambda$ is the optical frequency and $\eta = 0.87$ is the measured photodiode QE (very similar for both photodiodes).
The expression for $K$ is therefore
\begin{equation}
    K = \frac{R\, \eta \, q\, T}{h \nu \int_T h_{\textrm{norm}}(t)\, dt}
    = \SI{0.14}{\volt\per\milli\watt} \,,
\end{equation}
where $\int_T h_{\mathrm{norm}}(t)\, dt = \SI{2.57e-9}{\second}$, as obtained from the curves in Fig.~\ref{fig:h1_h2_shape_invariance}. The value of $K$ is consistent with the slopes obtained from the linear fits in Fig.~\ref{fig:h1_h2_linearity}, which are nearly identical, with $K_1 = K_2 = \SI{0.135}{\volt\per\milli\watt}$. The coefficients of determination are $R_1^2 = 0.9999$ and $R_2^2 = 0.9994$, confirming the excellent linearity of the detector response.

\subsection{Signal variance}

We now turn to the experimental validation of Eq.~\ref{eq:y_varianza_bilanciata}. To establish a temporal reference for selecting signal samples, we simultaneously recorded the detector’s difference photocurrent and the laser pulse signal from the internal monitor photodiode of the laser source. Both signals were then processed using a dedicated LabVIEW program.

\begin{figure}
\centering
\includegraphics[width=\linewidth]{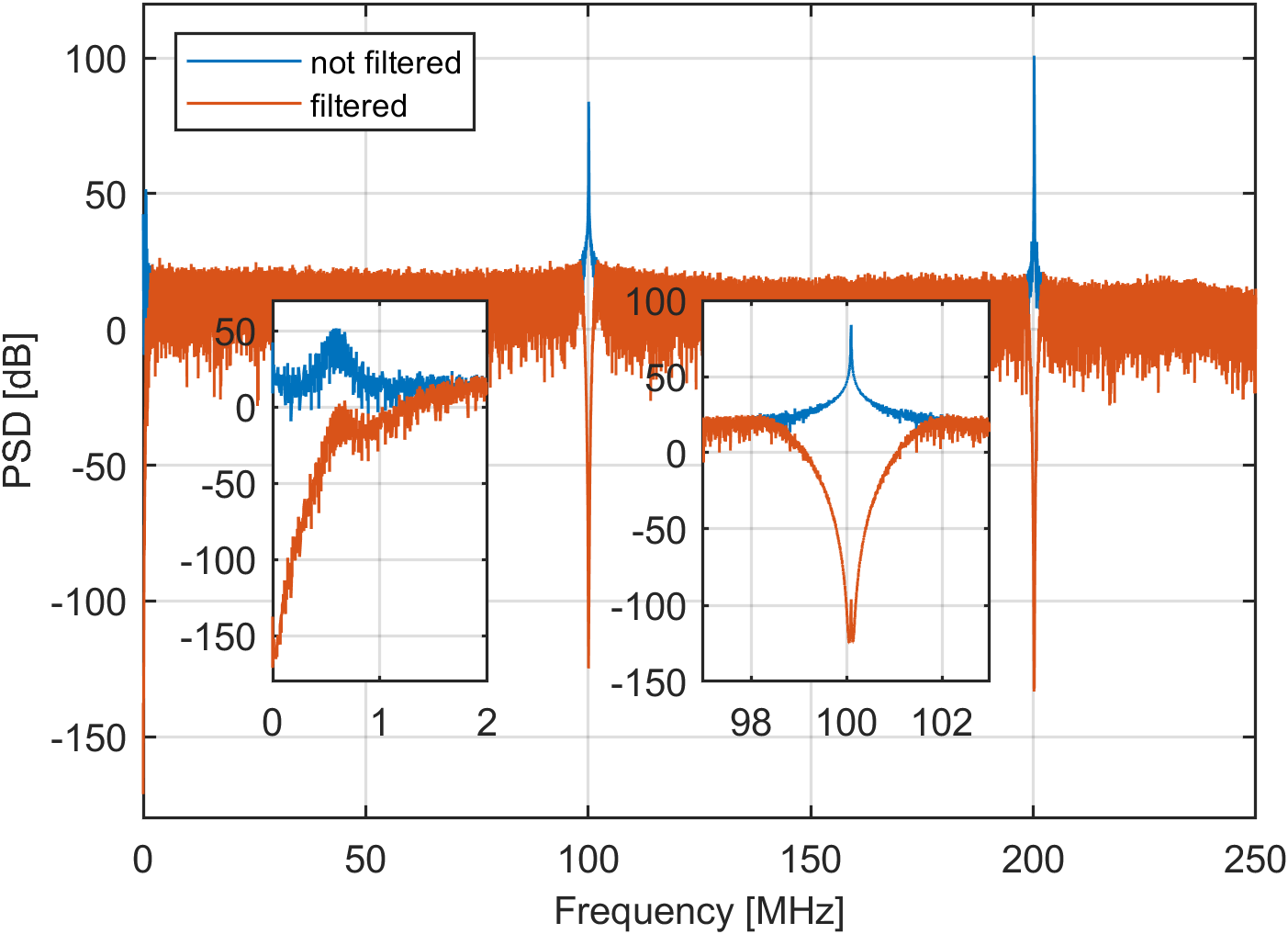}
\caption{Pulse train power spectrum before and after digital filtering.}
\label{fig:filtered_spectrum}
\end{figure}

To suppress low-frequency noise from the laser and amplifier, as well as laser noise at harmonics of the repetition rate, the acquired signals were digitally filtered within the acquisition program. For the low-frequency noise, we applied a sixth-order super-Gaussian high-pass filter with a \SI{3}{\mega\hertz} bandwidth. To remove noise at the source repetition rate harmonics, we used sixth-order super-Gaussian notch filters with a \SI{3}{\mega\hertz} FWHM and a spectral amplitude suppression of $10^{9}$ (see Figure~\ref{fig:filtered_spectrum}).

\begin{figure}
\centering
\includegraphics[width=0.8\linewidth]{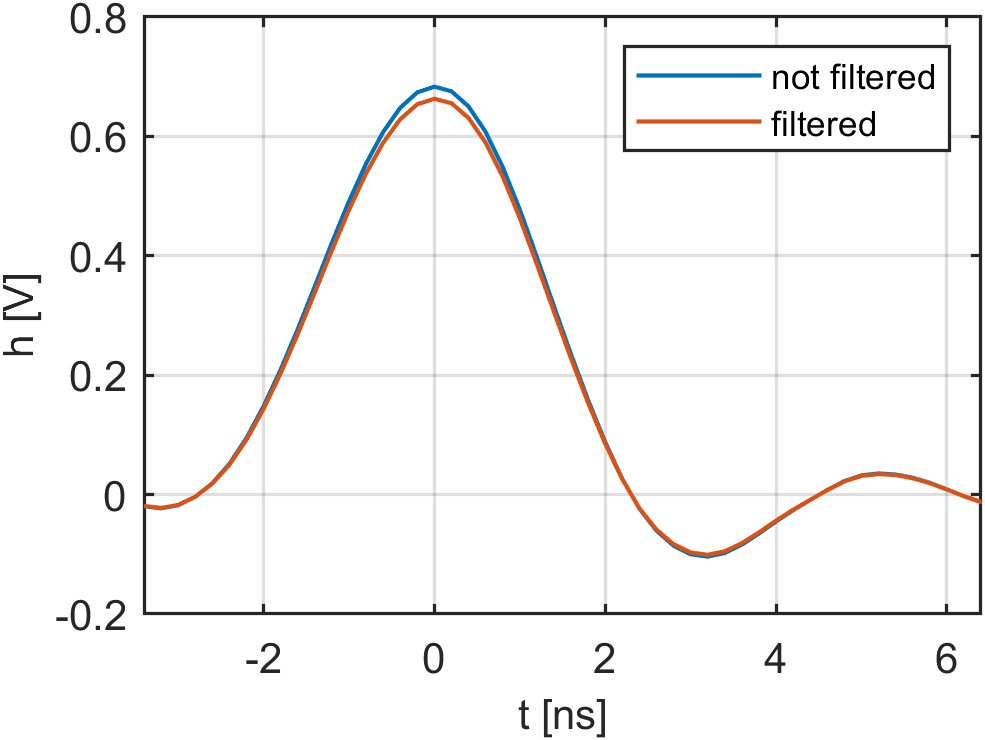}
\caption{Pulse shape before and after digital filtering.}
\label{fig:filtered_pulse}
\end{figure}

The signal $y(t)$ obtained after filtering, in response to a single pulse with photon number $n$, can now be written as
\begin{equation}
    y(t) = G \, F \, \tilde{y}(t) = n \, \frac{G \,F \,R \, \eta \, q}{\int_T h_{\textrm{norm}}(t)\, dt} \, h_{\textrm{norm}}(t) \,,
\end{equation}
where $G = 32.3 \pm 0.6$ is the voltage gain of the now reintroduced amplifier, measured to be constant over the relevant bandwidth, and $F = 0.971$ accounts for the effect of the notch filters. Over a single period $T$ of the source, the latter can in fact be approximated as a simple scaling of the single-pulse response, as shown in Figure~\ref{fig:filtered_pulse} (to obtain this, we started from the non-amplified pulses, constructed a time trace of the same length as those used for Figure~\ref{fig:filtered_spectrum}, retaining only the first pulse, performed the FFT, applied the filter, and finally transformed back to the time domain).

Equation~\ref{eq:y_varianza_bilanciata} can thus be rewritten as
\begin{equation}
    \Var{y(t)} = 2 \left( \frac{P T}{h \nu} \right) \left[ \frac{G \,F \,R \, \eta \, q}{\int_T h_{\textrm{norm}}(t)\, dt} \right]^2 h_{\textrm{norm}}^2(t)  \,+ \Mean{r^2(t)} \,.  
\end{equation}

\begin{figure}
\centering
\includegraphics[width=0.8\linewidth]{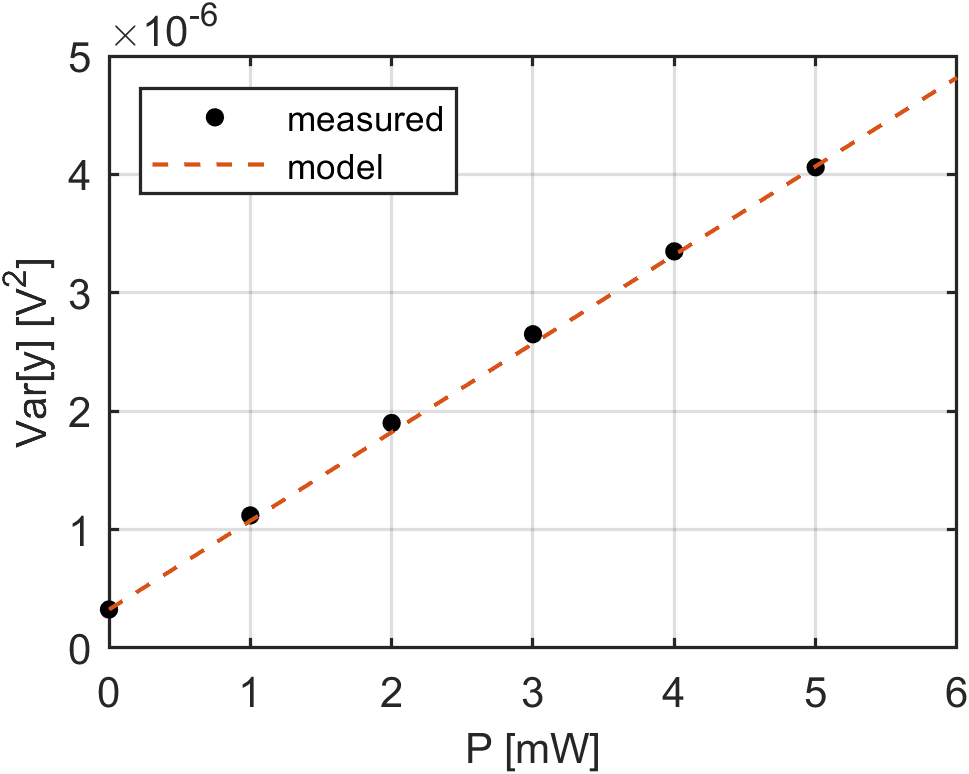}
\caption{Variance of the homodyne signal as a function of the optical power $P$, showing a linear dependence. The finite offset is due to electronic noise.}
\label{fig:variance_vs_P}
\end{figure}

\begin{figure}
\centering
\includegraphics[width=0.8\linewidth]{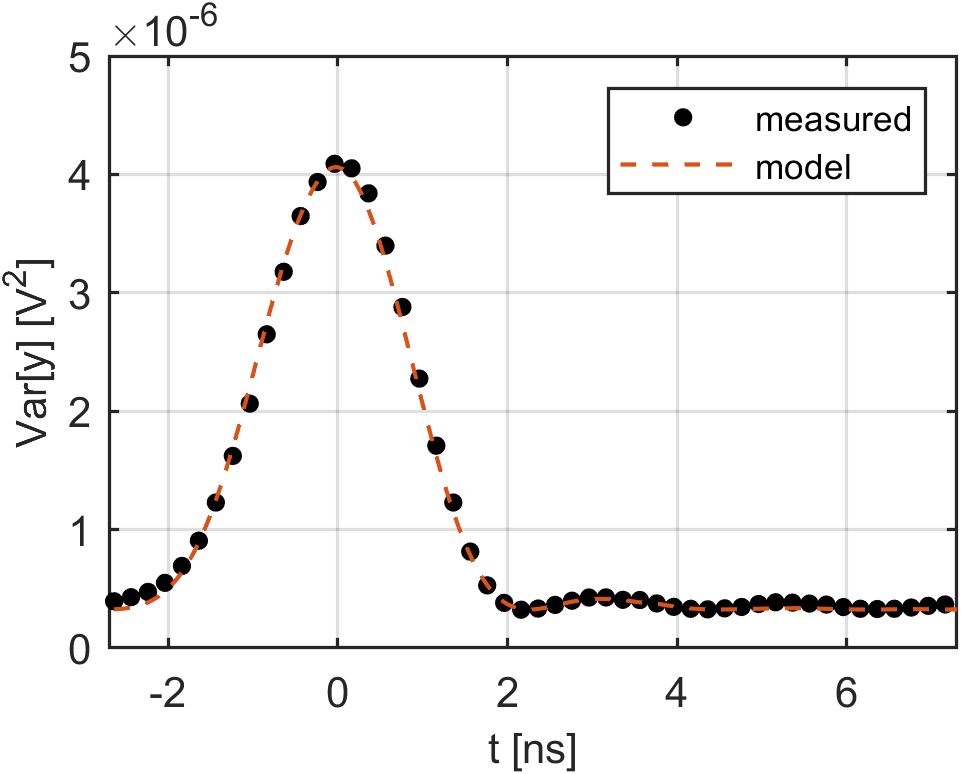}
\caption{Variance of the homodyne signal as a function of the delay with respect to the temporal reference. Apart from an electronic-noise offset, the variance reproduces the square of the temporal profile of the response function $h(t)$. The measurement was performed at an optical power of \SI{5}{\milli\watt}.}
\label{fig:variance_vs_t}
\end{figure}

In the measurement reported in Fig.~\ref{fig:variance_vs_P}, we show the linear dependence of the signal variance on the optical power. In our case, an offset of approximately \SI{0.3e-6}{\volt\squared} is present, arising from electronic noise. Figure~\ref{fig:variance_vs_t} instead shows the variance as a function of the delay with respect to the temporal reference - corresponding to the peak of the photocurrent - which reproduces, apart from the same offset, the profile of the square of the response function $h(t)$. This measurement was performed at an optical power of \SI{5}{\milli\watt}.

\begin{figure}
\centering
\includegraphics[width=\linewidth]{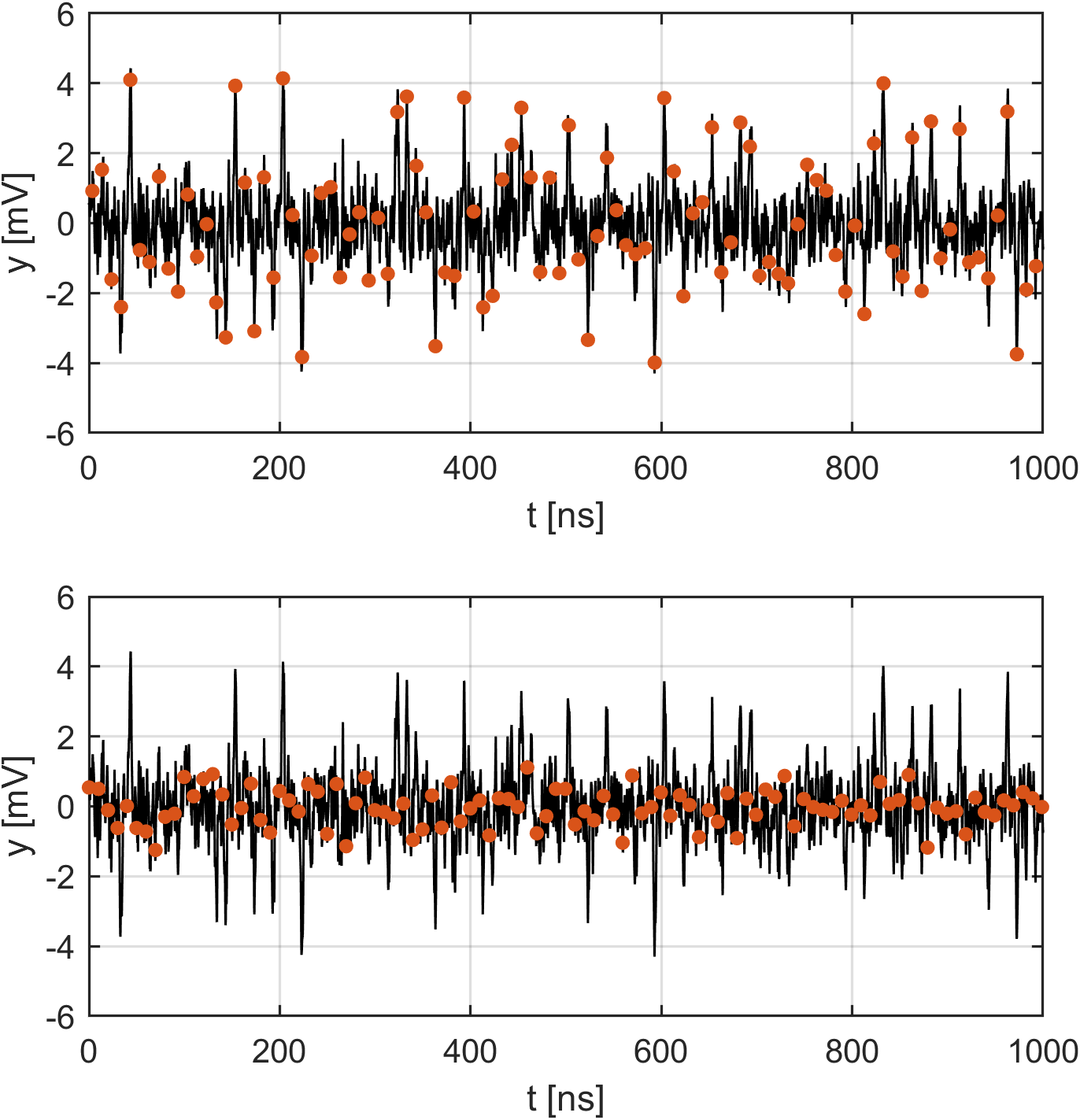}
\caption{Homodyne signal trace with samples selected at the photocurrent peak using the optimal delay time (top), and outside the photocurrent peak using a non-optimal delay time (bottom), resulting in a reduced variance in the latter case.}
\label{fig:spread_vs_time}
\end{figure}

Two examples showing a portion of the homodyne trace and the selected samples used to obtain the previous plot are reported in Fig.~\ref{fig:spread_vs_time}. In the first example, the samples correspond to the peak of the photocurrent, while in the second they are taken about \SI{3}{\nano\second} away from the peak. It is clearly visible that the spread of the samples is very different in the two cases, indicating that the instant corresponding to the photocurrent peak is the optimal choice to achieve the highest sensitivity to the shot noise.

\subsection{Optimal SNR}

We now discuss how the contribution of the electronic background to the variance can be minimized by integrating the pulses over a temporal window $\Delta t$, namely by experimentally investigating Eq.~\ref{eq:SNR}. Again, we start by rewriting the expression in terms of the experimental parameters as
\begin{equation}
    \mathrm{SNR}^2(\Delta t) = 1 + 2 \left( \frac{P T}{h \nu} \right) \left[ \frac{G \,F \,R \, \eta \, q}{\int_T h_{\textrm{norm}}(t)\, dt} \right]^2 \frac{H_{\mathrm{norm}}^2(\Delta t)}{\Mean{R^2(\Delta t)}} \,,
\end{equation}
where $H_{\textrm{norm}} = \int_{\Delta t} h_{\textrm{norm}}(t)\, dt$.

\begin{figure}
\centering
\includegraphics[width=0.8\linewidth]{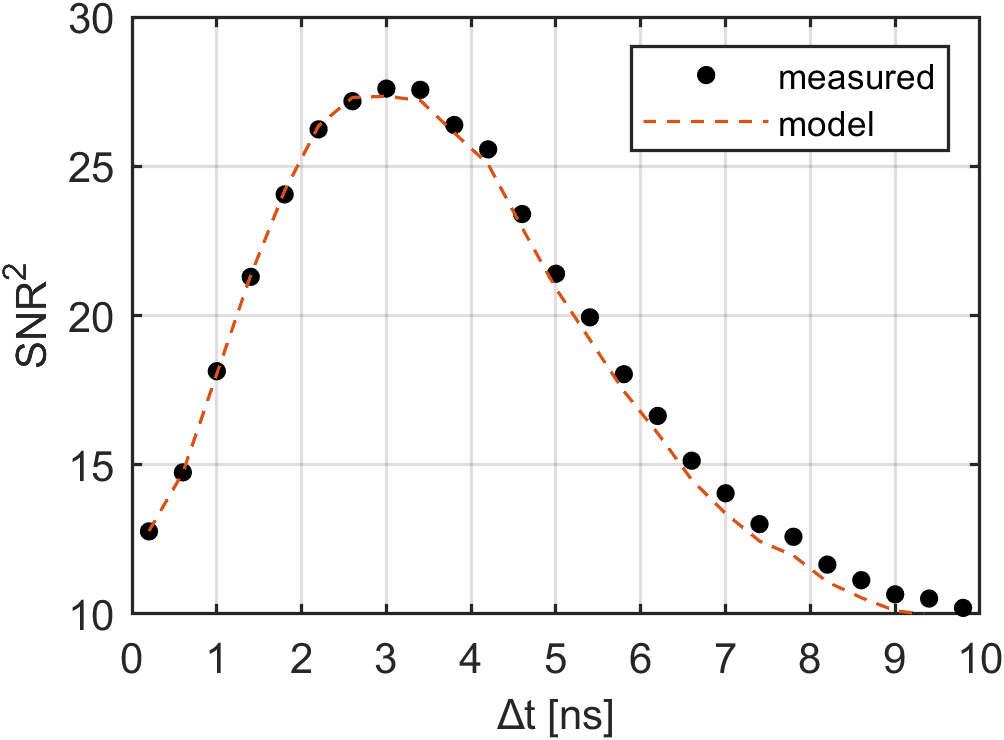}
\caption{Squared signal-to-noise ratio $\mathrm{SNR}^2$ as a function of the integration-window width $\Delta t$, centered on the voltage peak of the response function.}
\label{fig:SNR}
\end{figure}

Figure~\ref{fig:SNR} shows the behavior of $\mathrm{SNR}^2$ as a function of the integration-window width $\Delta t$ around the voltage peak. The maximum SNR is obtained for an integration window of approximately \SI{3}{\nano\second}, corresponding to a value of about \SI{14}{\deci\bel}, which is more than sufficient to observe quantum effects. At this SNR level, the shot noise clearly dominates over the electronic noise, allowing squeezing of the quadrature variance by several decibels to be reliably measured, with the only limitation in this case being the QE of the photodiodes.

\subsection{CMRR}

The CMRR of the detector can be optimized by adjusting the photodiode biases and fine-tuning the balance of the optical powers. This procedure is iterated until the signal peak at the laser repetition rate of \SI{100}{\mega\hertz} is minimized. In our case, we achieved a CMRR at \SI{100}{\mega\hertz} of \SI{58}{\deci\bel} at the maximum LO power of \SI{5}{\milli\watt}.

\begin{figure}
\centering
\includegraphics[width=0.8\linewidth]{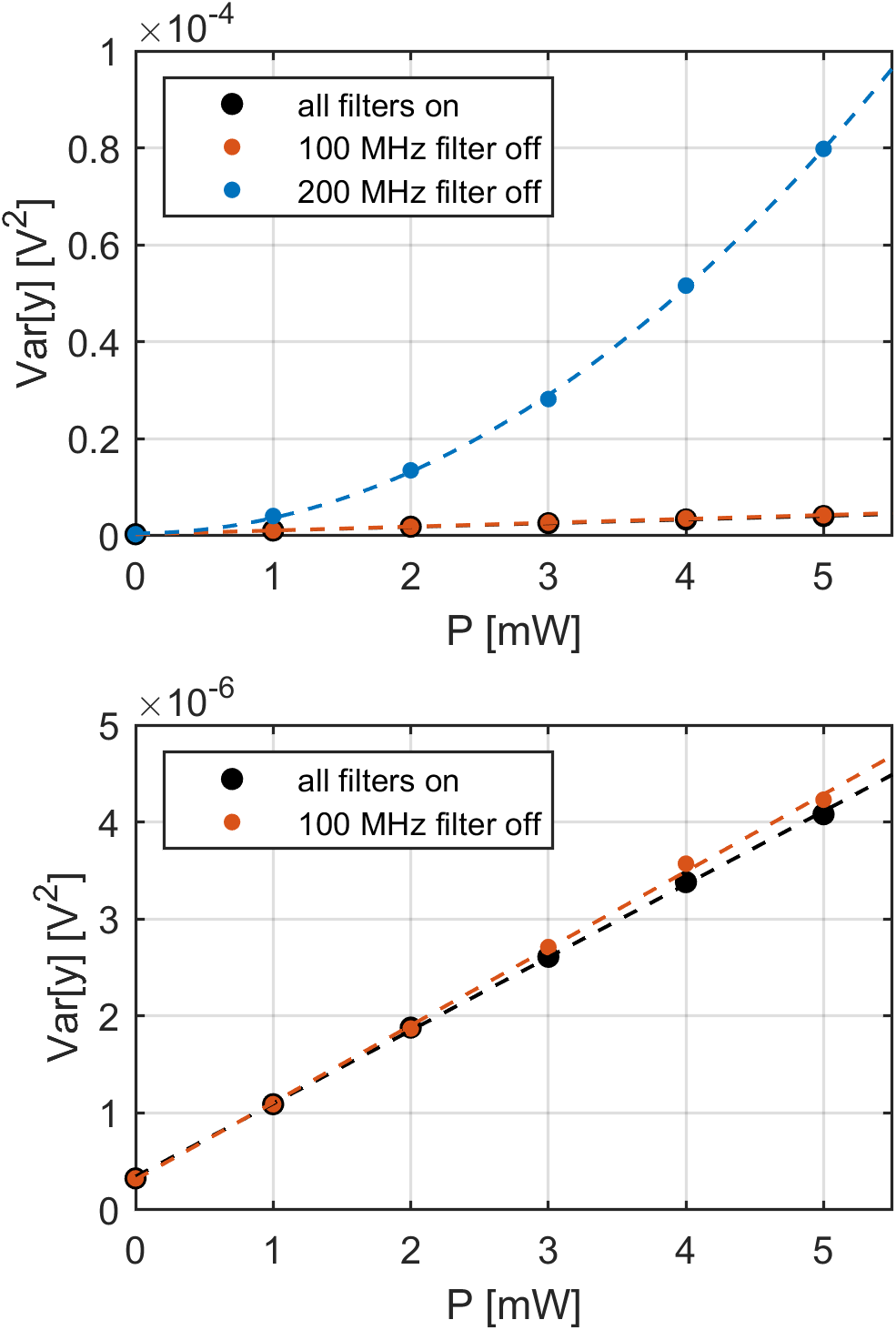}
\caption{On the top, variance of the detector output as a function of the LO power, measured with and without digital notch filters at \SI{100}{\mega\hertz} and \SI{200}{\mega\hertz}. On the bottom, a zoom better shows the effect of removing only the \SI{100}{\mega\hertz} filter. Since the CMRR is optimized at \SI{100}{\mega\hertz}, the \SI{200}{\mega\hertz} component still allows significant laser noise to pass. As a result, removing the \SI{200}{\mega\hertz} filter leads to a parabolic dependence (highlighted with a parabolic fit), whereas removing only the \SI{100}{\mega\hertz} filter preserves the linear behavior (highlighted with linear fits).}
\label{fig:variance_filtered_not_filtered}
\end{figure}

Figure~\ref{fig:variance_filtered_not_filtered} shows the variance as a function of the LO power, measured with and without the digital notch filters at \SI{100}{\mega\hertz} and \SI{200}{\mega\hertz}. It can be seen that the laser noise at \SI{100}{\mega\hertz} is suppressed, while it remains at \SI{200}{\mega\hertz}, leading to a quadratic dependence of the variance when the \SI{200}{\mega\hertz} filter is removed.

\subsection{Correlations}

\begin{figure}
\centering
\includegraphics[width=\linewidth]{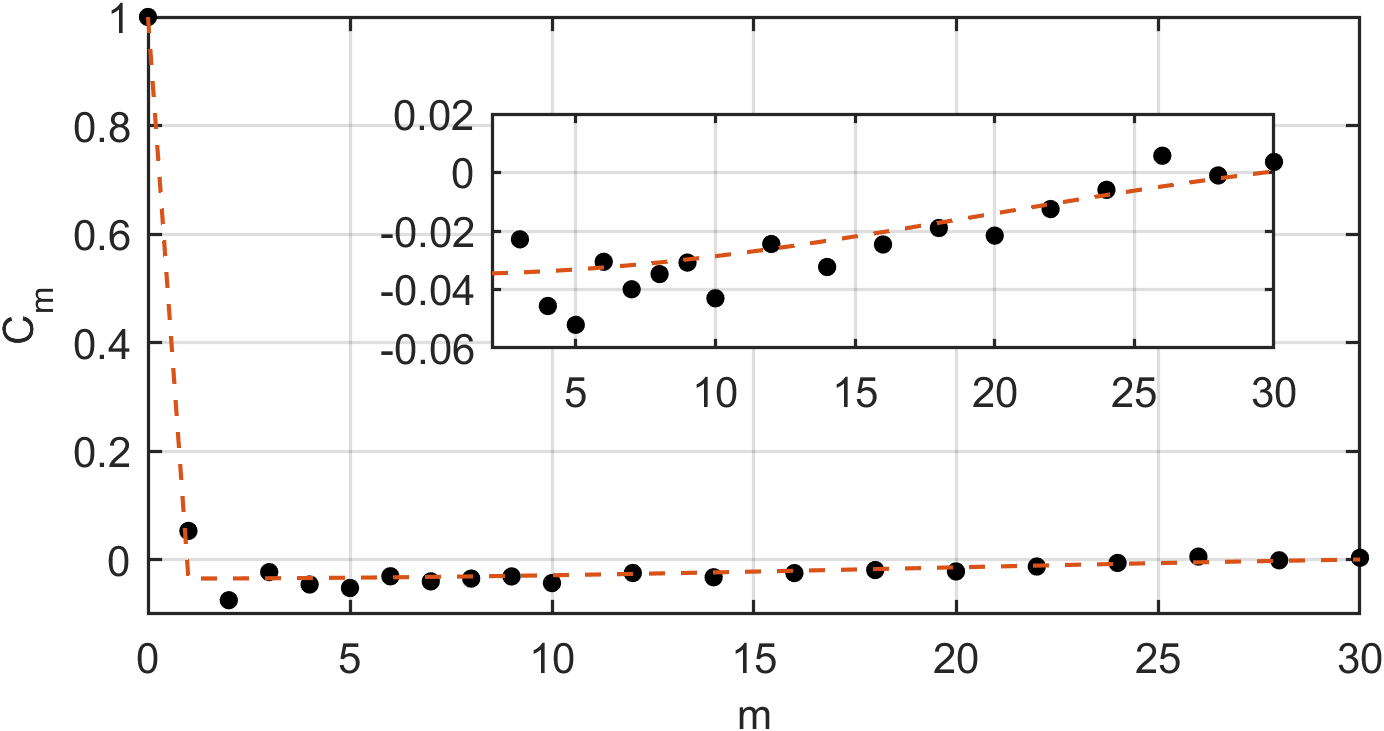}
\caption{Measured correlation coefficient $C_m$ between the signal sampled at the photocurrent peak of a given pulse and that of subsequent pulses. The correlations are negligible, indicating that the detector operates in the single-pulse regime. The dashed line represents the correlation trend calculated using Eq.~\ref{eq:corr_approx}.}
\label{fig:correlations}
\end{figure}

\begin{figure}
\centering
\includegraphics[width=\linewidth]{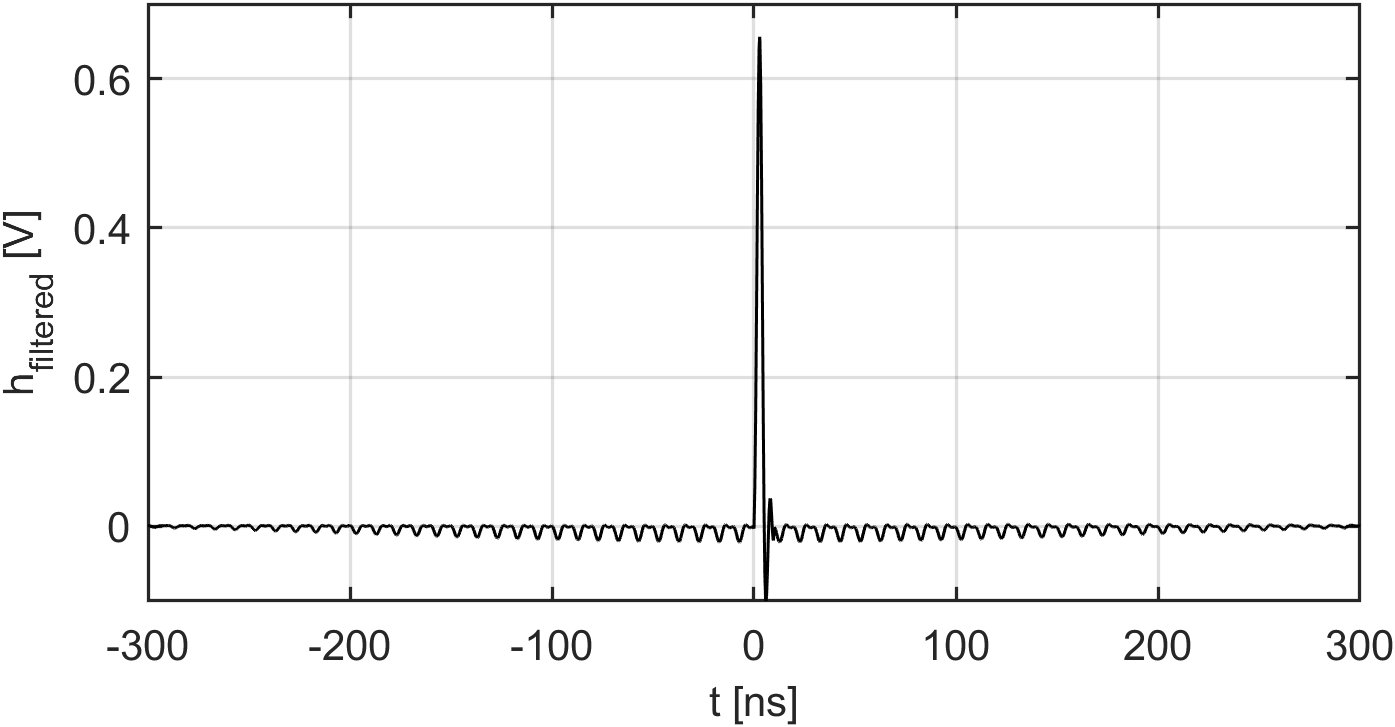}
\caption{Single-pulse response function after filtering with digital notch filters. The formation of a long tail, although limited in amplitude, explains the correlations measured between different pulses, as shown in Fig.~\ref{fig:correlations}.}
\label{fig:h_for_correlations}
\end{figure}

Finally, we present the measured correlations between different pulses.
The results shown in Fig.~\ref{fig:correlations} indicate that the correlation between a given pulse and all subsequent pulses is negligible, confirming that the detector can resolve and measure individual pulses. A closer inspection reveals small residual correlations that are negative and decay over a timescale of approximately $30$ pulses. The simple model of Eq.~\ref{eq:corr_approx} correctly reproduces the trend of these correlations when the single-pulse response function, filtered with the digital notch filters (see Fig.~\ref{fig:h_for_correlations}), is employed.

We can therefore attribute the residual inter-pulse correlations to the action of the notch filters. The amplitude of the negative peaks in Fig.~\ref{fig:h_for_correlations} increases, while their number decreases, as the bandwidth of the notch filters is increased. Consequently, the notch filters must be chosen sufficiently narrow in order to minimize correlations between neighboring pulses.

\section{Conclusions}

In this work, we have presented a balanced homodyne detector architecture specifically designed for pulsed optical sources operating at high repetition rates (\SI{100}{\mega\hertz}). By avoiding transimpedance amplifiers and instead directly amplifying the difference photocurrent extracted at the common node of the photodiodes without critical feedback loops, the detector effectively circumvents the nonlinearities and dynamic instabilities that typically arise in conventional TIA-based designs. A theoretical framework describing the detector response, noise characteristics, and pulse-to-pulse correlations has been developed and experimentally validated.

The experimental characterization confirms the excellent linearity of the detector response and the shot-noise-limited scaling of the signal variance with optical power. The optimal integration window of the homodyne trace yields a signal-to-noise ratio of approximately \SI{14}{\deci\bel}, sufficient for the observation of quantum effects such as quadrature squeezing. Moreover, the measured correlations between successive pulses are negligible, demonstrating that the detector operates effectively in the single-pulse regime. The simplicity and robustness of the proposed design make it a practical solution for high-repetition-rate homodyne measurements and a promising tool for experiments in pulsed quantum optics and continuous-variable quantum information processing.

\section*{Acknowledgements}
This work has been supported by INFN CSN5 within the project T4QC.

\bibliography{bibliography}

\end{document}